\documentclass[onecolumn,authoryear]{els-mrw}
\usepackage{amsmath,amssymb,amsfonts,amsthm,makeidx,graphicx}
\usepackage{bm}
\usepackage{hyperref}
\usepackage{CJKutf8}

\def \aj {{\rm {AJ}}}

\def \apj {{\rm {ApJ}}}

\def \apjs {{\rm {ApJS}}}

\def \aap {{\rm {A\&A}}}
\def \pasp{\rm {PASP}}

\def \apjl{\rm {ApJL}}

\def \mias{\,$\mu$as}

\begin{document}

\chapter{Astrometric detection of exoplanets}\label{chap1}

\author[1,2]{Fabo Feng}
\address[1]{\orgname{Tsung-Dao Lee Institute}, \orgdiv{Shanghai Jiao Tong University}, \orgaddress{Shengrong Road 520, Shanghai, 201210, People's Republic Of China}}
\address[2]{\orgname{School of Physics and Astronomy}, \orgdiv{Shanghai Jiao Tong University}, \orgaddress{800 Dongchuan Road, Shanghai, 200240, People's Republic of China}}

\articletag{This is an update of F. Feng, Astrometric detection of exoplanets, Editor(s):  Dimitri Veras, Snehil Sharma and Rajeswari R}

\maketitle

\textcolor{cyan}{
  \hrule
  \vspace{5pt}
\contentsline {section}{\numberline {1}Introduction}{1}{}
\contentsline {section}{\numberline {2}Types of astrometry}{2}{}
\contentsline {section}{\numberline {3}Detecting exoplanets with absolute astrometry}{2}{}
\contentsline {subsection}{\numberline {3.1}2D astrometry}{2}{}
\contentsline {subsection}{\numberline {3.2}1D astrometry}{4}{}
\contentsline {section}{\numberline {4}Detecting exoplanets with relative astrometry}{5}{}
\contentsline {subsection}{\numberline {4.1}Planet unresolved}{5}{}
\contentsline {subsection}{\numberline {4.2}Planet resolved}{6}{}
\contentsline {section}{\numberline {5}Sensitivity of astrometric exoplanet detection}{6}{}
\contentsline {section}{\numberline {6}Conclusions}{7}{}
\vspace{5pt}
\hrule
}


\begin{abstract}[Abstract]
As the most ancient branch of astronomy, astrometry has been developed for thousands of years. However, it has only recently become possible to utilize astrometry for the detection of exoplanets. Gaia, an astrometric surveyor of 1 billion stars, is capable of measuring the position of stars with a precision as high as 20\mias. Gaia is expected to discover more than 10,000 exoplanets by the end of its mission, surpassing the productivity of most exoplanet surveys.

In this chapter, I will introduce different techniques used to achieve
high-precision astrometry. Subsequently, I will explore how both
relative and absolute astrometry can be employed to detect
exoplanets. Finally, I will present the detection limit of the Gaia
astrometric survey.
\end{abstract}

\begin{BoxTypeA}[chap25:box2]{Key points}
\begin{itemize}
\item{\bf Astrometry:} The precise measurement of stars' positions and their movements across the celestial sphere.
\item {\bf Exoplanet:} A planet orbiting a star other than the Sun.
\item {\bf Reflex motion:} The movement of a star around the center of mass of the entire system, comprising the star and its associated planets.
\item {\bf Parallax:} The apparent motion of a star caused by the observer's movement relative to the Sun.
\end{itemize}
\end{BoxTypeA}

\section{Introduction}\label{sec:intro}
\addcontentsline{toc}{section}{Introduction} 

\begin{CJK*}{UTF8}{gbsn}
Astrometry is the specialized field of astronomy dedicated to the
precise measurement of the positions and movements of celestial
bodies. Millennia ago, ancient civilizations, including the Greeks,
Babylonians, and Chinese, meticulously tracked the paths of planets
and stars across the night sky. This practice resulted in the creation
of early star maps, such as the catalog compiled by the Greek
astronomer Hipparcus around 135 BC \citep{gysembergh22} and the {\it
  Shi's Classic of Stars} (石氏星经) by Chinese astronomer Shi Shen (石申) in the 4th century BC \citep{ho00}.
\end{CJK*}

In the late 16th century, Tycho Brahe conducted groundbreaking astrometric observations of approximately 1000 stars and planets. Johannes Kepler later compiled and published these observations in 1627 as the Rudolphine Tables. Kepler, utilizing these planetary astrometric data, went on to formulate the laws of planetary motion. In the 18th century, Sir William Herschel made significant contributions by discovering thousands of double stars, while Friedrich Bessel achieved a groundbreaking milestone by measuring the parallax of 61 Cygni with 9.6\% precision. This marked the first reliable measurement of the distance to a star beyond the solar system. The 19th-century introduction of photography revolutionized astrometry, as photographic plates enabled astronomers to capture and measure star positions with greater accuracy than traditional visual observations.

In 1989, the European Space Agency's Hipparcos satellite was launched,
conducting astrometric measurements for over a hundred thousand stars
down to approximately 11th magnitude with unprecedented accuracy,
reaching milli-arcsecond (mas) precision \citep{perryman97}. However,
it was later surpassed by its successor, Gaia, which achieved even
greater precision by measuring the positions of one billion stars with
an accuracy as fine as 20 micro-arcsecond (\mias; \citealt{gaia16b}). Prior to Gaia, the detection and confirmation of exoplanets through astrometry were limited. An example is the determination of the mass of GJ 876 b, where astrometric data from the Hubble Space Telescope (HST) Fine Guidance Sensor (FGS) were analyzed \citep{benedict02}.

While the Gaia epoch data is currently pending release, researchers
have already made significant strides in understanding exoplanets and
substellar companions. Through the combined analysis of high-precision
radial velocity data, along with astrometric data from Gaia and
Hipparcos, the masses and orbital parameters of hundreds of exoplanets
have been determined with remarkable precision
\citep{snellen18,brandt19,kervella19, feng22}. Furthermore, Gaia Data
Release 3 (DR3; \citealt{gaia23}) unveiled 64 new exoplanet candidates detected through astrometry \citep{arenou23}. In alignment with predictions by \cite{perryman14}, Gaia is poised to revolutionize exoplanet discoveries, with expectations surpassing 10,000 newly detected exoplanets, heralding a new era in this field.

Gaia is primarily focused on sub-mas absolute astrometry, achieved
through the observation of stars within a reference frame constructed
by astrometric observations of compact extragalactic sources, such as
more than one million quasars \citep{sergei22gaia}. Additionally, instruments like the optical interferometer FGS/HST and the infrared interferometer GRAVITY, integral to the Very Large Telescope Interferometer (VLTI), are capable of obtaining relative astrometry with a precision comparable to Gaia's absolute astrometry. FGS/HST can measure positional changes
between stars at a precision of about 200\mias, while GRAVITY/VLTI excels in detecting finer changes, ranging from 10
to 100\mias. Notably, GRAVITY achieved a remarkable 50\mias precision, determining the relative position of the renowned planet $\beta$ Pic b and enabling accurate measurements of its mass and orbital parameters \citep{nowak20}.

\section{Types of astrometry}\label{sec:astrometry}
\addcontentsline{toc}{section}{Types of astrometry} 

Astrometry involves the measurement of the position, motion, and
parallax of celestial bodies. To quantify these parameters, a
reference frame is established and realized through observations of
distant objects, such as quasars. Surveys like Gaia and the
Very-long-baseline interferometry (VLBI; e.g., \citealt{charlot20}) contribute to the establishment of absolute reference frames, enabling
the determination of ``absolute astrometry'' (or global astrometry) for celestial bodies. In
contrast, facilities like FGS/HST and GRAVITY/VLTI focus on measuring
the relative motions between celestial bodies, known as ``relative
astrometry.'' In relative astrometry, reference stars, usually distant
stars, are chosen, and the target star's motion is measured relative
to them (e.g., \citealt{benedict17}).

Three primary astrometric techniques are imaging (e.g., HST Wide Field
Camera), interferometry (e.g., GRAVITY/VLT), and drift scan (e.g.,
Gaia). Imaging astrometry captures target images over multiple epochs,
exemplified by the Cerro Tololo InterAmerican Observatory Parallax
Investigation (CTIOPI; \citealt{jao03}), facilitating efficient
astrometric determination. Interferometric astrometry utilizes
interference patterns from telescopes or light paths, achieving higher
angular resolution and detailed observation of fine celestial
details. Unlike the other techniques, drift-scan astrometry does not
track stars; instead, it allows stars to move across the detector over
time, providing high precision astrometry directly linked to temporal information.

\section{Detecting exoplanets with absolute astrometry}\label{sec:absolute}
\addcontentsline{toc}{section}{Detecting exoplanets with absolute astrometry}

\subsection{2D astrometry}
\addcontentsline{toc}{subsection}{2D astrometry}

For a planet with mass $m_p$ orbiting a star with mass $m_s$, the reflex motion of the star relative to the mass center is given by:
\begin{equation}
  \bm{r}_s(t)=
  \begin{bmatrix}
    x(t)\\
    y(t)\\
    z(t)
  \end{bmatrix}
  =\frac{m_p}{m_p+m_s}a
  \begin{bmatrix}
    \cos{E(t)}-e\\
    \sqrt{1-e^2}\sin{E(t)}\\
    0
  \end{bmatrix}~,
  \label{eq:rs}
\end{equation}
where $a$ is the semi-major axis of the planet with respect to the
star, $E(t)$ is the eccentricity anomaly, and $e$ is the
eccentricity. Here, $m_p$ and $m_s$ are respectively the mass of the
planet and the host star, and $m_s$ is typically determined through
methods such as isochrone fitting or mass-luminosity relation. The
semi-major axis for the reflex motion is $a_r =\frac{m_p}{m_p+m_s}a$.

\begin{figure}[t]
\centering
\includegraphics[width=.5\textwidth]{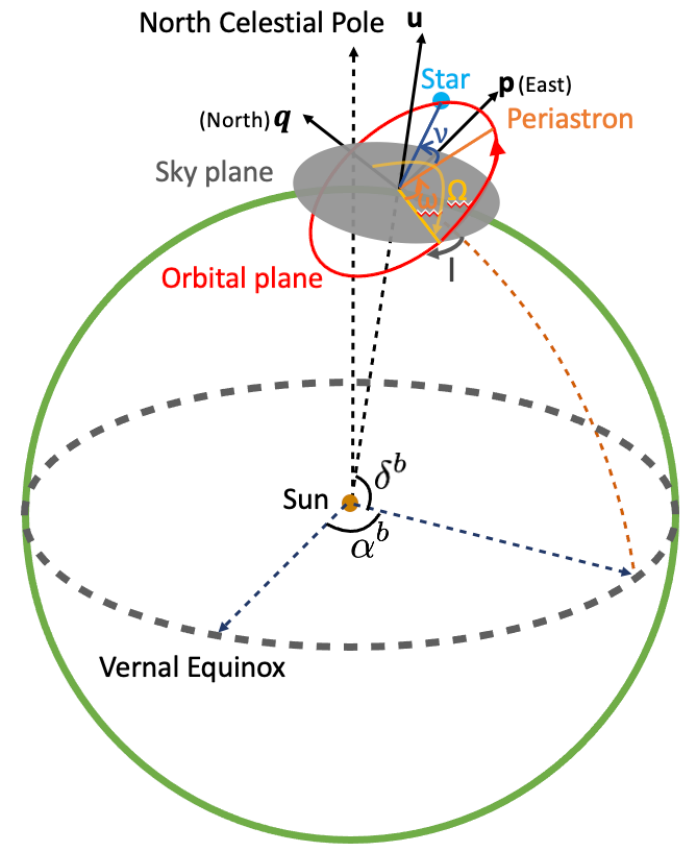}
\caption{Illustration of the coordinate system and orbital elements depicting stellar reflex motion (adapted from fig. 2 of \citealt{feng19pexo}). The true anomaly $\nu$ represents the angle measured from the periastron to the star's position. The longitude of ascending node $\Omega$ is the angle measured from the North to the ascending node. The inclination $I$ is the angle between the angular momentum of the orbital motion and $-\bm{u}$.}
\label{fig:elements}
\end{figure}

The stellar position $\bm{r}_s(t)$ is then converted to observer frame
(sky plane) coordinates, $\bm{r}_s^{\rm obs} (t)$, by applying Euler rotations using:
  \begin{equation}
    \bm{r}_s^{\rm obs}=R_z(\Omega)R_x(-I)R_z(\omega)\bm{r}_s(t)~,
    \label{eq:rotation}
  \end{equation}
where $\Omega$ is the longitude of ascending node, $\omega$ is the
argument of periastron, and $I$ is the inclination of the stellar
orbit with respect to the sky plane. It's important to note that the
longitude of ascending node and the argument of periastron of the
planetary (or secondary) motion should be $\Omega+\pi$ and
$\omega+\pi$, respectively. In the observer's frame, the directions of
the X axis (or $\bm q$, along North of the sky plane), Y axis (or $\bm p$,
along East of the sky plane), and Z axis (or $-\bm u$, along the
line of sight from the target to the observer) form a right-handed
Cartesian coordinate system in the observer frame. This coordinate
system corresponds to the first convention explained in the appendix of \cite{feng19pexo}. The orbital elements and coordinate systems are illustrated in Fig. \ref{fig:elements}.

\begin{figure}[t]
\centering
\includegraphics[width=1\textwidth]{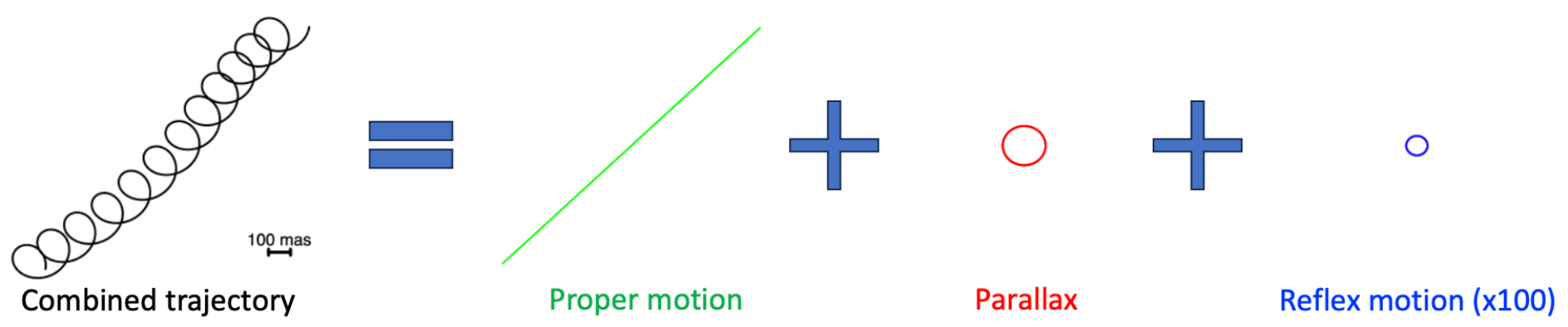}
\caption{Illustration showcasing different motions of a sun-like star
  when observed from a distance of 10\,pc. The reflex motion induced
  by a Jupiter analog is enhanced by 100 times for optimal
  visualization. Proper motions in the R.A. and decl. directions are set at 100\,mas.}
\label{fig:trajectory}
\end{figure}

  The expansion of
  eq. \ref{eq:rotation} yields the observed location of the star in
  the XYZ coordinate system:
  \begin{equation}
  \begin{bmatrix}
    X(t)\\
    Y(t)\\
    Z(t)\\
  \end{bmatrix}
  =
  \begin{bmatrix}
    A'&F'&-\sin{\Omega}\sin{I}\\
    B'&G'&\cos{\Omega}\sin{I}\\
    -\sin{\omega}\sin{I}&-\cos{\omega}\sin{I}&\cos{I}
  \end{bmatrix}
    \begin{bmatrix}
    x(t)\\
    y(t)\\
    0
  \end{bmatrix}
  ~,
\end{equation}
where
\begin{align}
  A'&=\cos{\Omega}\cos{\omega}-\sin{\Omega}\sin{\omega}\cos{I}\\
  B'&=\sin{\Omega}\cos{\omega}+\cos{\Omega}\sin{\omega}\cos{I}\\
  F'&=-\cos{\Omega}\sin{\omega}-\sin{\Omega}\cos{\omega}\cos{I}\\
  G'&=-\sin{\Omega}\sin{\omega}+\cos{\Omega}\cos{\omega}\cos{I}
\end{align}
are the scaled Thiele-Innes constants \citep{thiele83}, functions of inclination $I$, argument of periastron of the target star $\omega$, and longitude of ascending node $\Omega$. Multiplying $A', B', F', G'$ by the semi-major axis of the reflex motion $a_r$ defines the Thiele-Innes constants $A, B, F, G$. This assumes equivalence between the reflex motion and photocentric motion, as the planet's contribution is negligible in moving the photocenter away from the host star. The decomposition of the stellar trajectory into proper motion, parallax, and reflex motion is illustrated in Fig. \ref{fig:trajectory}.

Hence, the reflex motion of the star in the directions of Right Ascension (R.A. or $\alpha$) and Declination (decl. or $\delta$) induced by a planet is given by:
\begin{align}
\Delta\alpha^r_*(t)&=Y(t)/d=Bx(t)+Gy(t)~,\label{eq:reflex}\\\nonumber
\Delta\delta^r(t)&=X(t)/d=Ax(t)+Fy(t)~,
\end{align}
where $\Delta\alpha_*\equiv \Delta\alpha\cos\delta$ represents
R.A. offset projected onto the sky plane, $d$ is the heliocentric
distance of the star. For multiple-planet systems, the total reflex motion is the sum of the reflex motion due to individual planets, assuming no N-body interaction between planets.

In addition to the reflex motion, the barycenter of the star-planet system is described by:
\begin{align}
  \alpha_*^b&=\alpha_{*\rm ref}^b+\mu_\alpha^b(t-t_{\rm ref})+p_\alpha\varpi^b ~,\label{eq:bary}
\\\nonumber
  \delta^b&=\delta^b_{\rm ref}+\mu_\delta^b(t-t_{\rm ref})+p_\delta\varpi^b ~,
\end{align}
where $\bm{\kappa}^b\equiv(\alpha_*^b,\delta^b,\varpi^b,\mu_\alpha^b,\mu_\delta^b)$
represents the 5-parameter barycentric astrometry at epoch $t$, and
$p_\alpha$ and $p_\delta$ respectively represent the parallax factors
in the R.A. and decl. directions. However, this model does not account for effects such as perspective acceleration and gravitational lensing (e.g., \citealt{klioner92}). As these effects are typically calculated a priori, they are subtracted from the raw data during a calibration procedure before conducting subsequent model fitting.

The complete astrometric model for a target is given by:
\begin{align}
  \hat\alpha_*&=\alpha_*^b+\Delta\alpha_*^r~,\label{eq:full2D}\\\nonumber
  \hat\delta&=\delta^b+\Delta\delta^r~.
\end{align}
Typically, R.A. and decl. are measured relative to a reference position. Thus, the model and observation of the relative stellar position are represented as:
\begin{align}
  \hat{\bm{\zeta}}&=(\Delta\hat\alpha_*,\Delta\hat\delta)^T=(\hat{\alpha}_*-\alpha_{*\rm ref},\hat{\delta}-\delta_{\rm ref})^T~,\\\nonumber
  \bm{\zeta}&=(\Delta\alpha_*,\Delta\delta)^T=(\alpha_*-\alpha_{*\rm ref},\delta-\delta_{\rm ref})^T.
\end{align}

When both R.A. and decl. of a star are measured, and the astrometric noise is Gaussian, the likelihood is defined as:
\begin{equation}
 \mathcal{L}_{\rm 2D}=\prod_i^{N_{\rm epoch}}[(2\pi)^2|\Sigma_i|]^{-\frac{1}{2}}{\rm exp}\left\{-\frac{1}{2}[\hat{\bm{\zeta}}_i-\bm{\zeta}_i]^T\Sigma_i^{-1}[\hat{\bm{\zeta}}_i-\bm{\zeta}_i]\right\}~,
\label{eq:2Dlike}
\end{equation}
where $N_{\rm epoch}$ is the number of observation epochs, and
$\Sigma_i$ is the jitter-corrected covariance matrix of
$\bm{\zeta}_i$. Specifically, $\Sigma_i\equiv \Sigma_{\rm 0i}(1+J)$,
where $\Sigma_{\rm 0i}$ is the catalog covariance matrix for the
$i^{\rm th}$ epoch, and $J$ is referred to as ``relative astrometry
jitter.'' This relative jitter accounts for unknown systematics in
astrometric data.

\subsection{1D astrometry}
\addcontentsline{toc}{subsection}{1D astrometry}

\begin{figure}[t]
\centering
\includegraphics[width=.6\textwidth]{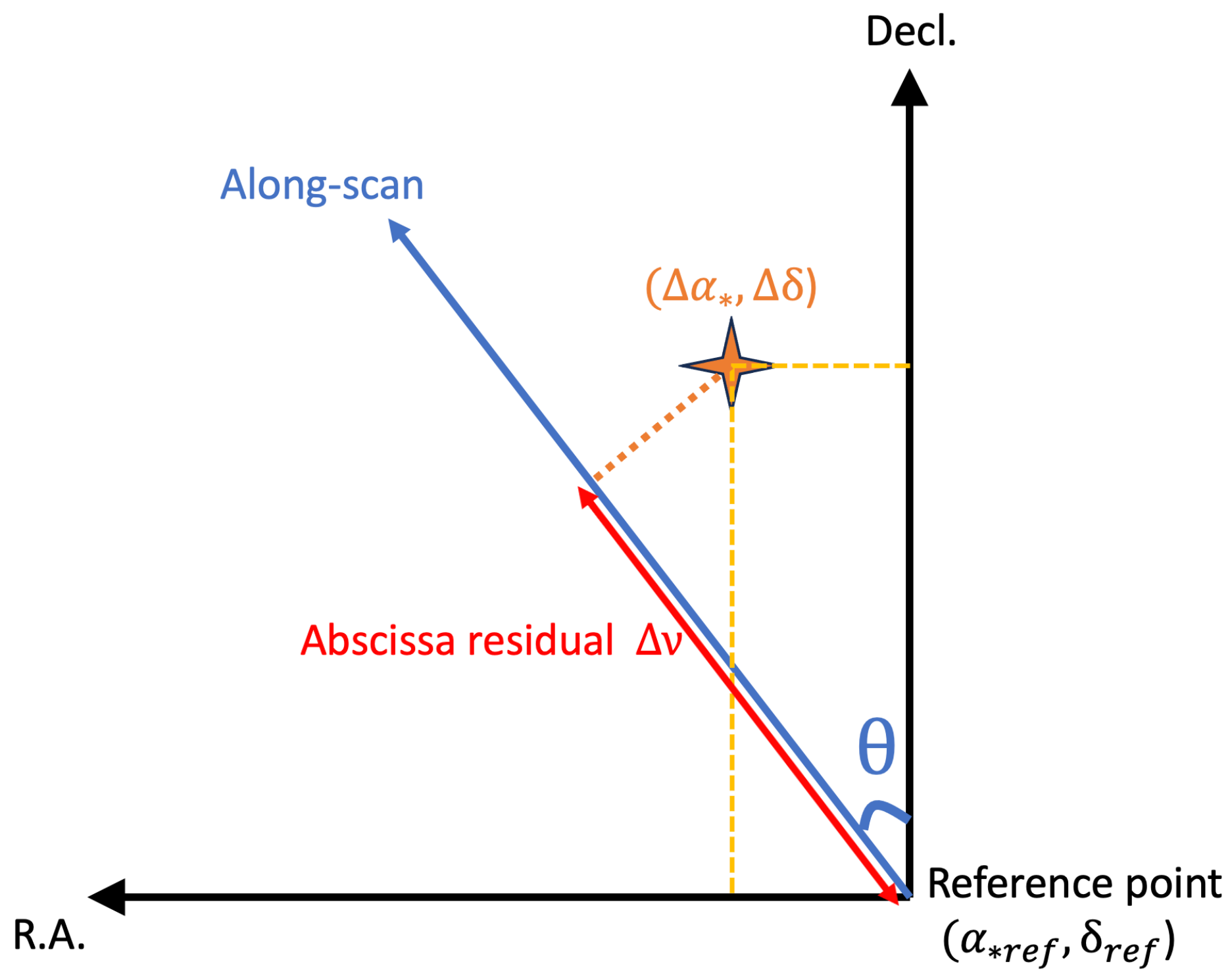}
\caption{Illustration of the conversion from R.A. and decl. to
  abscissa. }
\label{fig:abscissa}
\end{figure}

The drift-scan technique is commonly employed to enhance the
efficiency of astrometric surveys, exemplified by the successful
applications in the Hipparcos and Gaia missions. In this technique,
the along-scan (AL) coordinate is approximately one order of magnitude
more precise than the coordinate in the across-scan direction. The AL
coordinate, referred to as the ``abscissa,'' is derived from the 2D
model in eq. \ref{eq:full2D} by projecting the R.A. and
decl. onto the AL direction (as depicted in Fig. \ref{fig:abscissa}):
\begin{equation}
  \hat\nu_i=\hat\alpha_{*i}\sin\theta_i+\hat\delta_i\cos\theta_i~,\\
  \label{eq:xi}
\end{equation}
where $\theta$ represents the scan angle\footnote{This definition of scan angle is consistent with Gaia, while $\psi=\pi/2-\theta$ is the scan angle defined by Hipparcos.}.

Typically, the abscissae are defined as AL coordinates relative to the reference abscissa. Therefore, it is actually the abscissa residual. For a reference astrometry $\bm{\kappa}_{\rm ref}=(\alpha_{*\rm
  ref},\delta_{\rm ref},\varpi_{\rm ref},\mu_{\alpha,\rm
  ref},\mu_{\delta,\rm ref})^T$ at the reference epoch $t_{\rm ref}$, the abscissa residual is modeled as 
\begin{align}
  \Delta\hat\nu_i&= [(\alpha_*^b+\Delta\alpha_{*i}^r-\alpha_{*\rm
                     ref})+(\mu_\alpha^b-\mu_{\alpha,\rm
                     ref})(t_i-t_{\rm ref})]\sin\theta_i\nonumber\\
  &+[(\delta^b+\Delta\delta_i^r-\delta_{\rm ref})+(\mu_\delta^b-\mu_{\delta,\rm
                     ref})(t_i-t_{\rm ref})]\cos\theta_i\\
  &+(\varpi^b-\varpi_{\rm ref})p^{\rm AL}_i,\nonumber
\end{align}
where $\bm{\kappa}^b=(\alpha_{*}^b,\delta^b,\varpi^b, \mu_\alpha^b,\mu_\delta^b)^T$ is
the barycentric astrometry at the reference epoch, and $p^{\rm AL}_i$ is the AL parallax factor at epoch $t_i$. 
Defining $\Delta t_i$ as $t_i-t_{\rm ref}$ and the astrometric offsets
$\Delta\bm{\kappa}\equiv
(\Delta\alpha_*^b,\Delta\delta^b,\Delta\varpi^b,\Delta\mu_\alpha^b,\Delta\mu_\delta^b)$
as $\bm{\kappa}^b-\bm{\kappa}_{\rm ref}$, the above equation becomes
\begin{equation}
  \Delta\hat\nu_i=
  (\Delta\alpha_*^b+\Delta\alpha_{*i}^r+\Delta\mu_\alpha^b\Delta t_i)\sin\theta_i+(\Delta\delta^b+\Delta\delta_i^r+\Delta\mu_\delta^b\Delta t_i)\cos\theta_i +\Delta\varpi^bp^{\rm AL}_i.
\end{equation}
The corresponding likelihood for the above absissa model is
\begin{equation}
 \mathcal{L}_{\rm 1D}=\prod_i^{N_{\rm epoch}}[2\pi(\sigma_i^2+\sigma_J^2)]^{-\frac{1}{2}}{\rm exp}\left[-\frac{(\Delta\nu_i-\Delta\hat\nu_i)^2}{2(\sigma_i^2+\sigma_J^2)}\right]~,
\label{eq:1Dlike}
\end{equation}
where $\sigma_J$ is the jitter of abscissa, $\sigma_i$ is the error of
abscissa residual $\Delta\nu_i$.

\section{Detecting exoplanets with relative astrometry}\label{sec:relative}
\addcontentsline{toc}{section}{Detecting exoplanets with relative astrometry}

\subsection{Planet unresolved}
\addcontentsline{toc}{subsection}{Planet unresolved}

For a planet that is not resolved by imaging or interferometry, the
reflex motion of its host star can be measured relative to distant
reference stars in images taken at multiple epochs. All images (or
plates) are adjusted to align with the R.A. and decl. directions using
astrometric data from Gaia or Hipparcos\footnote{In practice, the
  alignment has a small uncertainty. This uncertainty introduces
  second-order modeling uncertainty, which can be minimized through
  iterative optimization of plate constants and astrometric parameters
  of reference stars.}. Additionally, the calibration for ground-based astrometry requires addressing the differential color refraction caused by atmospheric refraction \citep{jao03}. In FGS-like astrometry, the so-called ``lateral color'' resulting from the use of refractive optics is sometimes considered \citep{benedict99}. After calibration, the transformation of plates (or images) into a common frame, also known as a trail plate or constraint plate, is necessary.

\begin{figure}[t]
\centering
\includegraphics[width=.9\textwidth]{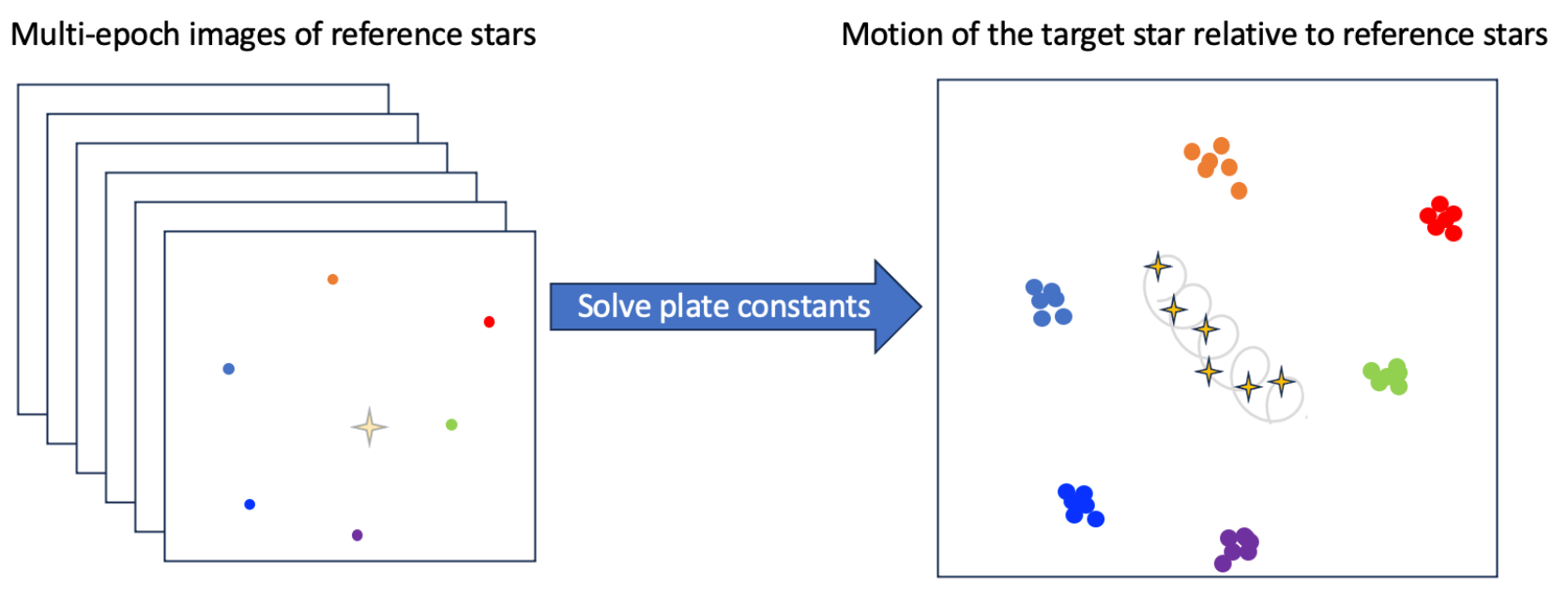}
\caption{Illustration of the relative astrometry for a star with
  an unresolved planet. The target star is not included in the process
of plate solution.}
\label{fig:plate}
\end{figure}

For a reference star $j$ in plate $i$ with coordinates $(x_{ij}, y_{ij})$, its coordinates in the constraint plate at epoch $t_{\rm ref}$ are given by:
\begin{align}
  \xi_{ij}&=a_ix_{ij}+b_iy_{ij}+c_{i}-\mu_{\alpha,j}(t_i-t_{\rm ref})-p_{\alpha,j}\varpi_j,\\\nonumber
  \eta_{ij}&=d_ix_{ij}+e_iy_{ij}+f_{i}-\mu_{\delta,j}(t_i-t_{\rm ref})-p_{\delta,j}\varpi_j,~
\end{align}
where $\bm{\beta}^p=(a_i, b_i, c_i, d_i, e_i, f_i)^T$ are referred to as ``plate constants,'' $t_i$ is the time when plate $i$ is obtained. Assuming that the average relative parallax and proper motions are zero, the plate constants and astrometric parameters of reference stars are iteratively optimized, typically using the GaussFit algorithm developed by \cite{jefferys88}. The principle of plate solution is illustrated in Fig. \ref{fig:plate}.

Once the plate parameters and astrometry of reference stars are
solved, the parameters for the target star (represented by index 0)
are determined by fixing these plate parameters at their optimal
values. The coordinates of the target star in the constraint plate are
\begin{align}
  \xi_i\equiv\xi_{i0}&=a_ix_{i0}+b_iy_{i0}+c_{i},\\\nonumber
  \eta_i\equiv\eta_{i0}&=d_ix_{i0}+e_iy_{i0}+f_{i}.~
\end{align}
The model for $\xi_i$ and $\eta_i$ is
\begin{align}
  \hat\xi_i&=\xi_0+\mu_{\alpha,0}(t_i-t_{\rm ref})+p_{\alpha,0}\varpi_0+\Delta\alpha^r_i,\\\nonumber
  \hat\eta_i&=\eta_0+\mu_{\delta,0}(t_i-t_{\rm ref})+p_{\delta,0}\varpi_0+\Delta\delta^r_i,~
\end{align}
where $(\xi_0,\eta_0)$ is the reference coordinate of the target star
in the constraint plate, $\Delta\alpha^r_i$ and
  $\Delta\delta^r_i$ are given in eq. \ref{eq:reflex}. The corresponding likelihood is given by:
\begin{equation}
   \mathcal{L}_{\rm unres}=\prod_i^{N_{\rm epoch}}[(2\pi)^2|\Sigma_i|]^{-\frac{1}{2}}{\rm exp}\left\{-\frac{1}{2}(\xi_i-\hat\xi_i,\eta_i-\hat\eta_i)\Sigma_i^{-1}(\xi_i-\hat\xi_i,\eta_i-\hat\eta_i)^T\right\}~,
 \end{equation}
 where $\Sigma_i$ is the covariance of $\xi_i$ and $\eta_i$. To model excess noise, the
 relative jitter $J$ may be included to define the
 jitter-corrected covariance, $\Sigma_i\equiv \Sigma_{i0}(1+J)$,
 where $\Sigma_{i0}$ is the measured covariance.

 \subsection{Planet resolved}
 \addcontentsline{toc}{subsection}{Planet resolved}
 
When a planet is resolved through direct imaging (e.g., \citealt{miles23}) or interferometric imaging (e.g., \citealt{nowak20}), its host star serves as a reference star to derive the relative astrometry. The reflex motion in eq. \ref{eq:reflex} is converted to the planetary motion using:
\begin{align}
  \Delta\hat\alpha_{*i}^p&=-\frac{m_s+m_p}{m_p}\Delta\alpha_i^r~,\\\nonumber
  \Delta\hat\delta_i^p&=-\frac{m_s+m_p}{m_p}\Delta\delta_i^r~.
  \label{eq:planetary}
\end{align}

Defining $\bm{\zeta}_i^p\equiv(\Delta\alpha_{*i}^p,\Delta\delta_i^p)^T$,
and
$\bm{\hat\zeta}_i^p\equiv(\Delta\hat\alpha_{*i}^p,\Delta\hat\delta_i^p)^T$, the corresponding likelihood is
\begin{equation}
   \mathcal{L}_{\rm res}=\prod_i^{N_{\rm epoch}}[(2\pi)^2|\Sigma_i|]^{-\frac{1}{2}}{\rm exp}\left\{-\frac{1}{2}(\bm{\zeta}_i^p-\bm{\hat\zeta}_i^p)^T\Sigma_i^{-1}(\bm{\zeta}_i^p-\bm{\hat\zeta}_i^p)\right\}~,
 \end{equation}
 where $\Sigma_i$ is the covariance of $\bm{\zeta}_i^p$. If the separation $\rho_i$ and position angle
 $\theta$ are provided (i.e., $\bm{\zeta}_i^p=(\rho_i,\theta_i)^T$),
 the model, $\bm{\hat\zeta}_i^p=(\hat\rho_i,\hat\theta_i)^T$, is
 given by\footnote{The atan2(y, x) function calculates the angle, in radians, between the positive x-axis and the ray extending from the origin to a point in the Cartesian plane.}:
  \begin{align}
    \hat\rho_i&=\sqrt{(\Delta\hat\alpha_{*i}^p)^2+(\Delta\hat\delta_i^p)^2}~,\label{eq:convert}\\
    \hat\theta_i&={\rm atan2}(\Delta\hat\alpha_{*i}^p,\Delta\hat\delta_i^p)~.  
  \end{align}

\section{Detection limit of Gaia}\label{sec:yield}
\addcontentsline{toc}{section}{Detection limit of Gaia}

\begin{figure}[t]
\centering
\includegraphics[width=.9\textwidth]{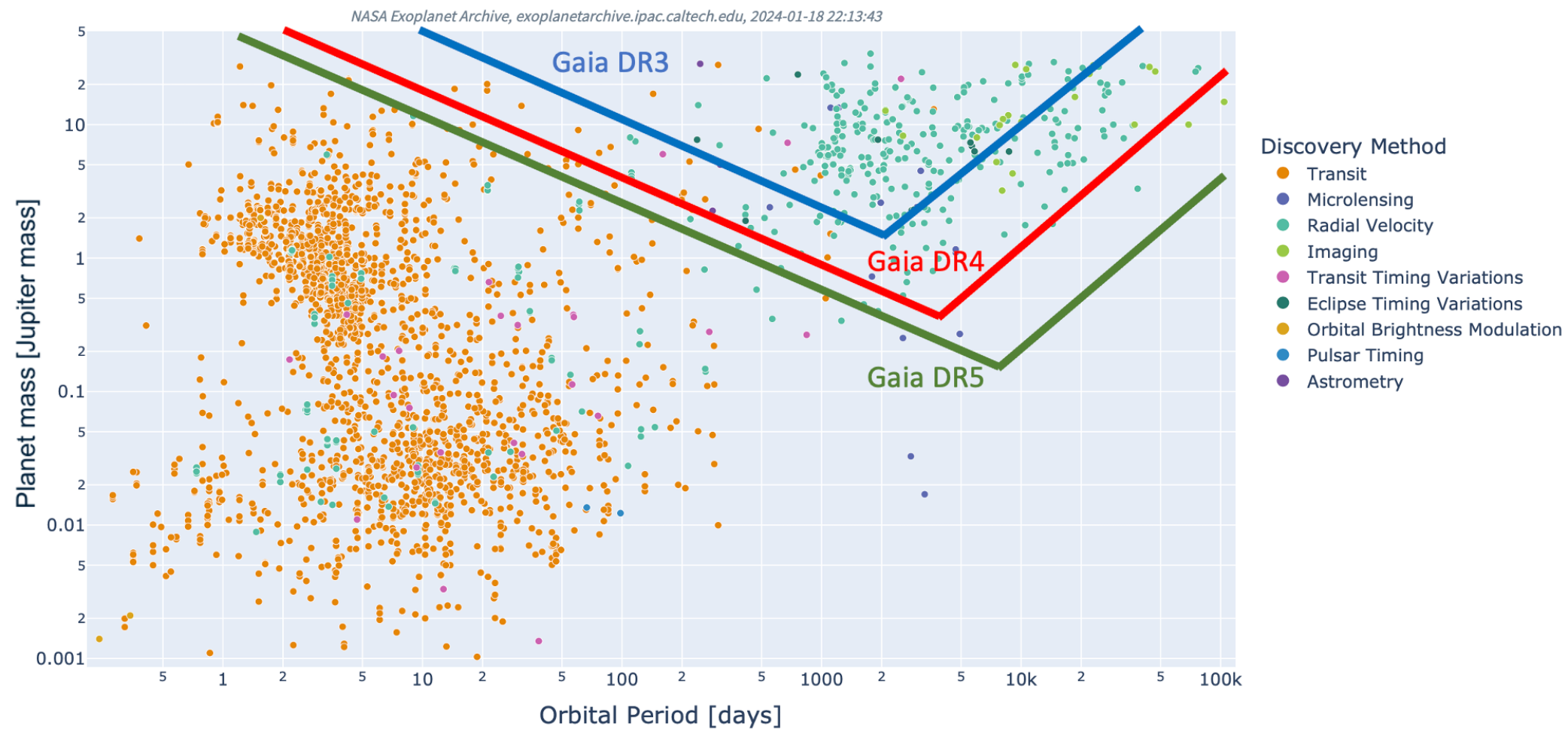}
\caption{Distribution of the currently detected exoplanets over
  orbital period and absolute planet mass. The mass for planets with
  orbital periods longer than 1000 days, discovered by radial
  velocity, is typically determined through combined analyses of
  radial velocity data and astrometric data from Gaia and
  Hipparcos. The detection limits of various Gaia data releases for a
  Sun-like star at a distance of 10\,pc are presented by lines with different colors. It's important to
    note that the sample comprises known exoplanets obtained from
    \href{exoplanetarchive.ipac.caltech.edu}{NASA Exoplanet Archive},
    and none of the exoplanets displayed in this figure were newly discovered by Gaia.}
\label{fig:yield}
\end{figure}

By measuring the astrometry for one billion stars to a precision as
high as 20\mias, Gaia brings a new era of exoplanet detection. The
expected sensitivity of Gaia and exoplanets with known absolute masses
are shown in Fig. \ref{fig:yield}. To estimate the sensitivity region
of Gaia in the m-a diagram, the astrometric signature of a reflex motion is given by:
\begin{equation}
  \alpha_{\rm astro}=\left(\frac{m_p}{m_s}\right)\left(\frac{a_p}{1~{\rm
      au}}\right)\left(\frac{d}{1~{\rm pc}}\right)^{-1}~{\rm arcsec}~.
\end{equation}
When the observation baseline $T$ is less
than half of the orbital period $P$, the signature is
\begin{equation}
  \alpha_{\rm astro}=\left(\frac{m_p}{m_s}\right)\left(\frac{a_p}{1~{\rm
      au}}\right)\left(\frac{d}{1~{\rm pc}}\right)^{-1}\left\{1-\sin\left[\pi\left(\frac{1}{2}-\frac{T}{P}\right)\right]\right\}~{\rm arcsec}~.
\end{equation}

The signal to noise ratio (SNR) is defined as 
\begin{equation}
  {\rm SNR}\equiv \alpha_{\rm astro}/\sigma_{\rm fov}~,
\end{equation}
where $\sigma_{\rm fov}$ is the along-scan accuracy per field of view
crossing. Following \cite{perryman14}, $\sigma_{\rm fov}$ can be
approximately derived from the precision of parallax $\sigma_\varpi$ following
\begin{equation}
  \sigma_{\rm fov} \approx 3.2\sigma_\varpi~.
\end{equation}
Assuming SNR$>5$ as the detection threshold and adopting precisions of
30, 10, and 7\mias$~$for the third, fourth, and fifth Gaia data
releases (DR3, DR4, and DR5) respectively, as
reported by \cite{gaia23} for bright stars, the detection limits of
different Gaia data releases for a Sun-like star at a distance of 10\,pc are shown in
Fig. \ref{fig:yield}. While Gaia DR3 can barely detect Jupiter
analogs, Gaia DR4 and DR5 are sensitive to Jupiter and even Saturn
analogs. According to \cite{perryman14}, Gaia would eventually detect
more than 10,000 exoplanets.

Moreover, the astrometry detection method
is sensitive to nearby exoplanets on wide orbits ($a>1$\,au or
$P>1$\,yr). This regime overlaps with the sensitivity ranges of both radial velocity and direct imaging methods. As a result, these techniques are commonly integrated to identify exoplanets characterized by exceptionally broad orbits and orbital periods spanning decades.

\section{Conclusions}\label{sec:conclusion}
\addcontentsline{toc}{section}{Conclusions}

In recent years, advancements in technology, such as drift-scan astrometry
and interferometry, have allowed for the measurement of stellar positions with sub-mas precision. With a precision in star position measurement as fine as 20\mias, Gaia is anticipated to identify over 10,000 exoplanets, predominantly cold giant planets, by the conclusion of its mission. This level of precision is approaching the
\mias$~$precision required for detecting nearby Earth-like planets around Sun-like stars. Looking ahead, the synergy between absolute and relative astrometry, such as combined analyses of FGS/HST and Gaia data, holds the potential to significantly enhance the observational baseline of astrometry. This improvement is crucial for detecting exoplanets, particularly those on extremely wide orbits.

\begin{ack}[Acknowledgments]

  This work is supported by Shanghai Jiao Tong University 2030
  Initiative. I express my gratitude to my students, Guangyao Xiao and
  Yifan Xuan, for providing valuable comments on the manuscript.
\end{ack}


\bibliographystyle{Harvard}
\bibliography{reference}

\begin{thebibliography*}{23}
\providecommand{\bibtype}[1]{}
\providecommand{\natexlab}[1]{#1}
{\catcode`\|=0\catcode`\#=12\catcode`\@=11\catcode`\\=12
|immediate|write|@auxout{\expandafter\ifx\csname
  natexlab\endcsname\relax\gdef\natexlab#1{#1}\fi}}
\renewcommand{\url}[1]{{\tt #1}}
\providecommand{\urlprefix}{URL }
\expandafter\ifx\csname urlstyle\endcsname\relax
  \providecommand{\doi}[1]{doi:\discretionary{}{}{}#1}\else
  \providecommand{\doi}{doi:\discretionary{}{}{}\begingroup
  \urlstyle{rm}\Url}\fi
\providecommand{\bibinfo}[2]{#2}
\providecommand{\eprint}[2][]{\url{#2}}

\bibtype{Article}%
\bibitem[{Benedict} et al.(1999)]{benedict99}
\bibinfo{author}{{Benedict} GF}, \bibinfo{author}{{McArthur} B},
  \bibinfo{author}{{Chappell} DW}, \bibinfo{author}{{Nelan} E},
  \bibinfo{author}{{Jefferys} WH}, \bibinfo{author}{{van Altena} W},
  \bibinfo{author}{{Lee} J}, \bibinfo{author}{{Cornell} D},
  \bibinfo{author}{{Shelus} PJ}, \bibinfo{author}{{Hemenway} PD},
  \bibinfo{author}{{Franz} OG}, \bibinfo{author}{{Wasserman} LH},
  \bibinfo{author}{{Duncombe} RL}, \bibinfo{author}{{Story} D},
  \bibinfo{author}{{Whipple} AL} and  \bibinfo{author}{{Fredrick} LW}
  (\bibinfo{year}{1999}), \bibinfo{month}{Aug.}
\bibinfo{title}{{Interferometric Astrometry of Proxima Centauri and Barnard's
  Star Using HUBBLE SPACE TELESCOPE Fine Guidance Sensor 3: Detection Limits
  for Substellar Companions}}.
\bibinfo{journal}{{\em \aj}} \bibinfo{volume}{118} (\bibinfo{number}{2}):
  \bibinfo{pages}{1086--1100}. \bibinfo{doi}{\doi{10.1086/300975}}.
\eprint{astro-ph/9905318}.

\bibtype{Article}%
\bibitem[{Benedict} et al.(2002)]{benedict02}
\bibinfo{author}{{Benedict} GF}, \bibinfo{author}{{McArthur} BE},
  \bibinfo{author}{{Fredrick} LW}, \bibinfo{author}{{Harrison} TE},
  \bibinfo{author}{{Slesnick} CL}, \bibinfo{author}{{Rhee} J},
  \bibinfo{author}{{Patterson} RJ}, \bibinfo{author}{{Skrutskie} MF},
  \bibinfo{author}{{Franz} OG}, \bibinfo{author}{{Wasserman} LH},
  \bibinfo{author}{{Jefferys} WH}, \bibinfo{author}{{Nelan} E},
  \bibinfo{author}{{van Altena} W}, \bibinfo{author}{{Shelus} PJ},
  \bibinfo{author}{{Hemenway} PD}, \bibinfo{author}{{Duncombe} RL},
  \bibinfo{author}{{Story} D}, \bibinfo{author}{{Whipple} AL} and
  \bibinfo{author}{{Bradley} AJ} (\bibinfo{year}{2002}), \bibinfo{month}{Sep.}
\bibinfo{title}{{Astrometry with the Hubble Space Telescope: A Parallax of the
  Fundamental Distance Calibrator {\ensuremath{\delta}} Cephei}}.
\bibinfo{journal}{{\em \aj}} \bibinfo{volume}{124} (\bibinfo{number}{3}):
  \bibinfo{pages}{1695--1705}. \bibinfo{doi}{\doi{10.1086/342014}}.
\eprint{astro-ph/0206214}.

\bibtype{Article}%
\bibitem[{Benedict} et al.(2017)]{benedict17}
\bibinfo{author}{{Benedict} GF}, \bibinfo{author}{{McArthur} BE},
  \bibinfo{author}{{Nelan} EP} and  \bibinfo{author}{{Harrison} TE}
  (\bibinfo{year}{2017}), \bibinfo{month}{Jan.}
\bibinfo{title}{{Astrometry with Hubble Space Telescope Fine Guidance
  Sensors{\textemdash}A Review}}.
\bibinfo{journal}{{\em \pasp}} \bibinfo{volume}{129} (\bibinfo{number}{971}):
  \bibinfo{pages}{012001}.
  \bibinfo{doi}{\doi{10.1088/1538-3873/129/971/012001}}.
\eprint{1610.05176}.

\bibtype{Article}%
\bibitem[{Brandt} et al.(2019)]{brandt19}
\bibinfo{author}{{Brandt} TD}, \bibinfo{author}{{Dupuy} TJ} and
  \bibinfo{author}{{Bowler} BP} (\bibinfo{year}{2019}), \bibinfo{month}{Oct.}
\bibinfo{title}{{Precise Dynamical Masses of Directly Imaged Companions from
  Relative Astrometry, Radial Velocities, and Hipparcos-Gaia DR2
  Accelerations}}.
\bibinfo{journal}{{\em \aj}} \bibinfo{volume}{158} (\bibinfo{number}{4}),
  \bibinfo{eid}{140}. \bibinfo{doi}{\doi{10.3847/1538-3881/ab04a8}}.
\eprint{1811.07285}.

\bibtype{Article}%
\bibitem[{Charlot} et al.(2020)]{charlot20}
\bibinfo{author}{{Charlot} P}, \bibinfo{author}{{Jacobs} CS},
  \bibinfo{author}{{Gordon} D}, \bibinfo{author}{{Lambert} S},
  \bibinfo{author}{{de Witt} A}, \bibinfo{author}{{B{\"o}hm} J},
  \bibinfo{author}{{Fey} AL}, \bibinfo{author}{{Heinkelmann} R},
  \bibinfo{author}{{Skurikhina} E}, \bibinfo{author}{{Titov} O},
  \bibinfo{author}{{Arias} EF}, \bibinfo{author}{{Bolotin} S},
  \bibinfo{author}{{Bourda} G}, \bibinfo{author}{{Ma} C},
  \bibinfo{author}{{Malkin} Z}, \bibinfo{author}{{Nothnagel} A},
  \bibinfo{author}{{Mayer} D}, \bibinfo{author}{{MacMillan} DS},
  \bibinfo{author}{{Nilsson} T} and  \bibinfo{author}{{Gaume} R}
  (\bibinfo{year}{2020}), \bibinfo{month}{Dec.}
\bibinfo{title}{{The third realization of the International Celestial Reference
  Frame by very long baseline interferometry}}.
\bibinfo{journal}{{\em \aap}} \bibinfo{volume}{644}, \bibinfo{eid}{A159}.
  \bibinfo{doi}{\doi{10.1051/0004-6361/202038368}}.
\eprint{2010.13625}.

\bibtype{Article}%
\bibitem[{Feng} et al.(2019)]{feng19pexo}
\bibinfo{author}{{Feng} F}, \bibinfo{author}{{Lisogorskyi} M},
  \bibinfo{author}{{Jones} HRA}, \bibinfo{author}{{Kopeikin} SM},
  \bibinfo{author}{{Butler} RP}, \bibinfo{author}{{Anglada-Escud{\'e}} G} and
  \bibinfo{author}{{Boss} AP} (\bibinfo{year}{2019}), \bibinfo{month}{Oct.}
\bibinfo{title}{{PEXO: A Global Modeling Framework for Nanosecond Timing,
  Microarcsecond Astrometry, and {\ensuremath{\mu}}m s$^{-1}$ Radial
  Velocities}}.
\bibinfo{journal}{{\em \apjs}} \bibinfo{volume}{244} (\bibinfo{number}{2}),
  \bibinfo{eid}{39}. \bibinfo{doi}{\doi{10.3847/1538-4365/ab40b6}}.
\eprint{1910.01750}.

\bibtype{Article}%
\bibitem[{Feng} et al.(2022)]{feng22}
\bibinfo{author}{{Feng} F}, \bibinfo{author}{{Butler} RP},
  \bibinfo{author}{{Vogt} SS}, \bibinfo{author}{{Clement} MS},
  \bibinfo{author}{{Tinney} CG}, \bibinfo{author}{{Cui} K},
  \bibinfo{author}{{Aizawa} M}, \bibinfo{author}{{Jones} HRA},
  \bibinfo{author}{{Bailey} J}, \bibinfo{author}{{Burt} J},
  \bibinfo{author}{{Carter} BD}, \bibinfo{author}{{Crane} JD},
  \bibinfo{author}{{Dotti} FF}, \bibinfo{author}{{Holden} B},
  \bibinfo{author}{{Ma} B}, \bibinfo{author}{{Ogihara} M},
  \bibinfo{author}{{Oppenheimer} R}, \bibinfo{author}{{O'Toole} SJ},
  \bibinfo{author}{{Shectman} SA}, \bibinfo{author}{{Wittenmyer} RA},
  \bibinfo{author}{{Wang} SX}, \bibinfo{author}{{Wright} DJ} and
  \bibinfo{author}{{Xuan} Y} (\bibinfo{year}{2022}), \bibinfo{month}{Sep.}
\bibinfo{title}{{3D Selection of 167 Substellar Companions to Nearby Stars}}.
\bibinfo{journal}{{\em \apjs}} \bibinfo{volume}{262} (\bibinfo{number}{1}),
  \bibinfo{eid}{21}. \bibinfo{doi}{\doi{10.3847/1538-4365/ac7e57}}.
\eprint{2208.12720}.

\bibtype{Article}%
\bibitem[{Gaia Collaboration} et al.(2016)]{gaia16b}
\bibinfo{author}{{Gaia Collaboration}}, \bibinfo{author}{{Prusti} T},
  \bibinfo{author}{{de Bruijne} JHJ}, \bibinfo{author}{{Brown} AGA} and
  \bibinfo{author}{et~al.} (\bibinfo{year}{2016}), \bibinfo{month}{Nov.}
\bibinfo{title}{{The Gaia mission}}.
\bibinfo{journal}{{\em \aap}} \bibinfo{volume}{595}, \bibinfo{eid}{A1}.
  \bibinfo{doi}{\doi{10.1051/0004-6361/201629272}}.
\eprint{1609.04153}.

\bibtype{Article}%
\bibitem[{Gaia Collaboration} et al.(2022)]{sergei22gaia}
\bibinfo{author}{{Gaia Collaboration}}, \bibinfo{author}{{Klioner} SA},
  \bibinfo{author}{{Lindegren} L}, \bibinfo{author}{{Mignard} F} and
  \bibinfo{author}{et~al.} (\bibinfo{year}{2022}), \bibinfo{month}{Nov.}
\bibinfo{title}{{Gaia Early Data Release 3. The celestial reference frame
  (Gaia-CRF3)}}.
\bibinfo{journal}{{\em \aap}} \bibinfo{volume}{667}, \bibinfo{eid}{A148}.
  \bibinfo{doi}{\doi{10.1051/0004-6361/202243483}}.
\eprint{2204.12574}.

\bibtype{Article}%
\bibitem[{Gaia Collaboration} et al.(2023{\natexlab{a}})]{arenou23}
\bibinfo{author}{{Gaia Collaboration}}, \bibinfo{author}{{Arenou} F},
  \bibinfo{author}{{Babusiaux} C}, \bibinfo{author}{{Barstow} MA},
  \bibinfo{author}{{Faigler} S}, \bibinfo{author}{{Jorissen} A} and
  \bibinfo{author}{et~al.} (\bibinfo{year}{2023}{\natexlab{a}}),
  \bibinfo{month}{Jun.}
\bibinfo{title}{{Gaia Data Release 3. Stellar multiplicity, a teaser for the
  hidden treasure}}.
\bibinfo{journal}{{\em \aap}} \bibinfo{volume}{674}, \bibinfo{eid}{A34}.
  \bibinfo{doi}{\doi{10.1051/0004-6361/202243782}}.
\eprint{2206.05595}.

\bibtype{Article}%
\bibitem[{Gaia Collaboration} et al.(2023{\natexlab{b}})]{gaia23}
\bibinfo{author}{{Gaia Collaboration}}, \bibinfo{author}{{Vallenari} A},
  \bibinfo{author}{{Brown} AGA}, \bibinfo{author}{{Prusti} T} and
  \bibinfo{author}{et~al.} (\bibinfo{year}{2023}{\natexlab{b}}),
  \bibinfo{month}{Jun.}
\bibinfo{title}{{Gaia Data Release 3. Summary of the content and survey
  properties}}.
\bibinfo{journal}{{\em \aap}} \bibinfo{volume}{674}, \bibinfo{eid}{A1}.
  \bibinfo{doi}{\doi{10.1051/0004-6361/202243940}}.
\eprint{2208.00211}.

\bibtype{Article}%
\bibitem[{GRAVITY Collaboration} et al.(2020)]{nowak20}
\bibinfo{author}{{GRAVITY Collaboration}}, \bibinfo{author}{{Nowak} M},
  \bibinfo{author}{{Lacour} S}, \bibinfo{author}{{Molli{\`e}re} P} and
  \bibinfo{author}{et~al.} (\bibinfo{year}{2020}), \bibinfo{month}{Jan.}
\bibinfo{title}{{Peering into the formation history of {\ensuremath{\beta}}
  Pictoris b with VLTI/GRAVITY long-baseline interferometry}}.
\bibinfo{journal}{{\em \aap}} \bibinfo{volume}{633}, \bibinfo{eid}{A110}.
  \bibinfo{doi}{\doi{10.1051/0004-6361/201936898}}.
\eprint{1912.04651}.

\bibtype{Article}%
\bibitem[Gysembergh et al.(2022)]{gysembergh22}
\bibinfo{author}{Gysembergh V}, \bibinfo{author}{J.~Williams P} and
  \bibinfo{author}{Zingg E} (\bibinfo{year}{2022}).
\bibinfo{title}{New evidence for hipparchus’ star catalogue revealed by
  multispectral imaging}.
\bibinfo{journal}{{\em Journal for the History of Astronomy}}
  \bibinfo{volume}{53} (\bibinfo{number}{4}): \bibinfo{pages}{383--393}.

\bibtype{Book}%
\bibitem[Ho(2000)]{ho00}
\bibinfo{author}{Ho PY} (\bibinfo{year}{2000}).
\bibinfo{title}{Li, Qi and Shu: An introduction to science and civilization in
  China}, \bibinfo{publisher}{Courier Corporation}.

\bibtype{Article}%
\bibitem[{Jao} et al.(2003)]{jao03}
\bibinfo{author}{{Jao} WC}, \bibinfo{author}{{Henry} TJ},
  \bibinfo{author}{{Subasavage} JP}, \bibinfo{author}{{Bean} JL},
  \bibinfo{author}{{Costa} E}, \bibinfo{author}{{Ianna} PA} and
  \bibinfo{author}{{M{\'e}ndez} RA} (\bibinfo{year}{2003}),
  \bibinfo{month}{Jan.}
\bibinfo{title}{{The Solar Neighborhood. VII. Discovery and Characterization of
  Nearby Multiples in the CTIO Parallax Investigation}}.
\bibinfo{journal}{{\em \aj}} \bibinfo{volume}{125} (\bibinfo{number}{1}):
  \bibinfo{pages}{332--342}. \bibinfo{doi}{\doi{10.1086/345515}}.

\bibtype{Article}%
\bibitem[{Jefferys} et al.(1988)]{jefferys88}
\bibinfo{author}{{Jefferys} WH}, \bibinfo{author}{{Fitzpatrick} MJ} and
  \bibinfo{author}{{McArthur} BE} (\bibinfo{year}{1988}), \bibinfo{month}{Jan.}
\bibinfo{title}{{GaussFit{\textemdash}A system for least squares and robust
  estimation}}.
\bibinfo{journal}{{\em Celestial Mechanics}} \bibinfo{volume}{41}
  (\bibinfo{number}{1-4}): \bibinfo{pages}{39--49}.
  \bibinfo{doi}{\doi{10.1007/BF01238750}}.

\bibtype{Article}%
\bibitem[{Kervella} et al.(2019)]{kervella19}
\bibinfo{author}{{Kervella} P}, \bibinfo{author}{{Arenou} F},
  \bibinfo{author}{{Mignard} F} and  \bibinfo{author}{{Th{\'e}venin} F}
  (\bibinfo{year}{2019}), \bibinfo{month}{Mar.}
\bibinfo{title}{{Stellar and substellar companions of nearby stars from Gaia
  DR2. Binarity from proper motion anomaly}}.
\bibinfo{journal}{{\em \aap}} \bibinfo{volume}{623}, \bibinfo{eid}{A72}.
  \bibinfo{doi}{\doi{10.1051/0004-6361/201834371}}.
\eprint{1811.08902}.

\bibtype{Article}%
\bibitem[{Klioner} and {Kopeikin}(1992)]{klioner92}
\bibinfo{author}{{Klioner} SA} and  \bibinfo{author}{{Kopeikin} SM}
  (\bibinfo{year}{1992}), \bibinfo{month}{Aug.}
\bibinfo{title}{{Microarcsecond Astrometry in Space: Relativistic Effects and
  Reduction of Observations}}.
\bibinfo{journal}{{\em \aj}} \bibinfo{volume}{104}: \bibinfo{pages}{897}.
  \bibinfo{doi}{\doi{10.1086/116284}}.

\bibtype{Article}%
\bibitem[{Miles} et al.(2023)]{miles23}
\bibinfo{author}{{Miles} BE}, \bibinfo{author}{{Biller} BA},
  \bibinfo{author}{{Patapis} P} and  \bibinfo{author}{et~al.}
  (\bibinfo{year}{2023}), \bibinfo{month}{Mar.}
\bibinfo{title}{{The JWST Early-release Science Program for Direct Observations
  of Exoplanetary Systems II: A 1 to 20 {\ensuremath{\mu}}m Spectrum of the
  Planetary-mass Companion VHS 1256-1257 b}}.
\bibinfo{journal}{{\em \apjl}} \bibinfo{volume}{946} (\bibinfo{number}{1}),
  \bibinfo{eid}{L6}. \bibinfo{doi}{\doi{10.3847/2041-8213/acb04a}}.
\eprint{2209.00620}.

\bibtype{Article}%
\bibitem[{Perryman} et al.(1997)]{perryman97}
\bibinfo{author}{{Perryman} MAC}, \bibinfo{author}{{Lindegren} L},
  \bibinfo{author}{{Kovalevsky} J}, \bibinfo{author}{{Hoeg} E},
  \bibinfo{author}{{Bastian} U}, \bibinfo{author}{{Bernacca} PL},
  \bibinfo{author}{{Cr{\'e}z{\'e}} M}, \bibinfo{author}{{Donati} F},
  \bibinfo{author}{{Grenon} M}, \bibinfo{author}{{Grewing} M},
  \bibinfo{author}{{van Leeuwen} F}, \bibinfo{author}{{van der Marel} H},
  \bibinfo{author}{{Mignard} F}, \bibinfo{author}{{Murray} CA},
  \bibinfo{author}{{Le Poole} RS}, \bibinfo{author}{{Schrijver} H},
  \bibinfo{author}{{Turon} C}, \bibinfo{author}{{Arenou} F},
  \bibinfo{author}{{Froeschl{\'e}} M} and  \bibinfo{author}{{Petersen} CS}
  (\bibinfo{year}{1997}), \bibinfo{month}{Jul.}
\bibinfo{title}{{The HIPPARCOS Catalogue}}.
\bibinfo{journal}{{\em \aap}} \bibinfo{volume}{323}: \bibinfo{pages}{L49--L52}.

\bibtype{Article}%
\bibitem[{Perryman} et al.(2014)]{perryman14}
\bibinfo{author}{{Perryman} M}, \bibinfo{author}{{Hartman} J},
  \bibinfo{author}{{Bakos} G{\'A}} and  \bibinfo{author}{{Lindegren} L}
  (\bibinfo{year}{2014}), \bibinfo{month}{Dec.}
\bibinfo{title}{{Astrometric Exoplanet Detection with Gaia}}.
\bibinfo{journal}{{\em \apj}} \bibinfo{volume}{797}, \bibinfo{eid}{14}.
  \bibinfo{doi}{\doi{10.1088/0004-637X/797/1/14}}.
\eprint{1411.1173}.

\bibtype{Article}%
\bibitem[{Snellen} and {Brown}(2018)]{snellen18}
\bibinfo{author}{{Snellen} IAG} and  \bibinfo{author}{{Brown} AGA}
  (\bibinfo{year}{2018}), \bibinfo{month}{Aug.}
\bibinfo{title}{{The mass of the young planet Beta Pictoris b through the
  astrometric motion of its host star}}.
\bibinfo{journal}{{\em Nature Astronomy}} \bibinfo{volume}{2}:
  \bibinfo{pages}{883--886}. \bibinfo{doi}{\doi{10.1038/s41550-018-0561-6}}.
\eprint{1808.06257}.

\bibtype{Article}%
\bibitem[{Thiele}(1883)]{thiele83}
\bibinfo{author}{{Thiele} TN} (\bibinfo{year}{1883}), \bibinfo{month}{Jan}.
\bibinfo{title}{{Neue Methode zur Berechung von Doppelsternbahnen}}.
\bibinfo{journal}{{\em Astronomische Nachrichten}} \bibinfo{volume}{104}:
  \bibinfo{pages}{245}.

\end{thebibliography*}

\end{document}